\documentclass[12pt]{article}
\usepackage{graphicx}
\usepackage{amssymb}
\usepackage{mathpazo}
\usepackage{epstopdf}
\usepackage{float}					
\newcommand{\ignore}[1]{}
\newcommand{\be}{\begin{equation}} \newcommand{\ee}{\end{equation}}
\newcommand{\ba}{\begin{eqnarray}} \newcommand{\ea}{\end{eqnarray}}
\newcommand{\nn}{\nonumber} 
\newcommand{\ra}{\rightarrow} 
\renewcommand{\d}{\mathrm{d}}

\renewcommand{\a}{\alpha} \renewcommand{\b}{\beta}

\newcommand{\p}{\partial}

\def\slasha#1{\setbox0=\hbox{$#1$}#1\hskip-\wd0\hbox to\wd0{\hss\sl/\/\hss}}
\def\slashb#1{\setbox0=\hbox{$#1$}#1\hskip-\wd0\dimen0=5pt\advance
       \dimen0 by-\ht0\advance\dimen0 by\dp0\lower0.5\dimen0\hbox
         to\wd0{\hss\sl/\/\hss}}

\def\ket#1{\left| #1\right>}

\date{}

\begin{document}


\title{A Pionic Hadron \\ Explains the \\ Muon Magnetic Moment Anomaly } 

\author{Rainer W.~Schiel and John P. Ralston \\ {\it Department of Physics and Astronomy } \\ { \it  University of Kansas,  Lawrence, KS 66045  } }

\maketitle

\begin{abstract}
A significant discrepancy exists between experiment and calculations of the muon's magnetic moment. We find that standard formulas for the hadronic vacuum polarization term have overlooked pionic states known to exist.  Coulomb binding alone guarantees $\pi^{+} \pi^{-}$ states that quantum mechanically mix with the $\rho$ meson.  A simple 2-state mixing model explains the magnetic moment discrepancy for a mixing angle of order $\a \sim 10^{-2}$. The relevant physical state is predicted to give a tiny observable bump in the ratio $R(\, s\, )$ of $e^{+}e^{-}$ annihilation at a low energy not previously searched.  The burden of proof is reversed for claims that conventional physics cannot explain the muon's anomalous moment. 

\end{abstract}

1. Calculations of the muon's magnetic moment do not currently agree with experiment.  The discrepancy is of order three standard deviations and quite important.  Among other things, uncertainties of the anomalous moment feed directly to precision tests of the Standard Model, including the Higgs mass, as well as providing primary constraints on new physics such as supersymmetry.  In terms of the Land\'e $g$ factor, the ``anomaly'' $a_{\mu} =(\, g-2\, )/2$ experimentally observed in muons has become the quintessential precision test of quantum electrodynamics (\, QED\, ).  The current world average for $a_\mu$ is \cite{BNL,Davier}:
\ba
a_\mu^{experimental} = (\, 11659208.0 \pm 6.3 \, )\times  10^{-10}. \nn 
\ea The Standard Model theoretical prediction \cite{theory} for $a_\mu$ is
\ba
a_\mu^{theory} = (\, 11659180.4 \pm 5.1\, )\times  10^{-10}. \label{disc}
\ea
The values include contributions from QED, weak interactions including $W$ and $Z$ exchanges, plus hadronic (\, strongly interacting\, ) effects.  Uncertainties treat the errors for the hadronic, light-by-light scattering and electroweak contributions in quadrature.  The comparison to 9 digit accuracy represents thousands of physicist-years of experimental and theoretical labor. 
 
We reviewed contributions considered the most problematic. The {\it hadronic vacuum polarization} term involves virtual photons emitted by the muon, which fluctuate into strongly interacting particles before being re-absorbed. The corresponding contribution $a_{\mu}^{had, \, LO}$ is expressed using the ratio $R(\, s\, ) = \sigma(\, e^{+}e^{-} ~\ra hadrons\, )/\sigma(\, e^{+}e^{-} \ra \mu^{+}\mu^{-}\, )$, where $\sigma$ represents the total cross section. Because $R(\, s\, )$ involves strong interactions it must be taken from experiments. The formula for $a_{\mu}^{had, \, LO}$ is \cite{Bouchiat}
\ba \label{dispint}
a_\mu^{had, LO} = \frac{\alpha^2}{3 \pi^2} \int_{s_{min}}^\infty \d s \frac{K(\, x(\, s\, )\, )}{s} R(\, s\, ), \label{form}
\ea
where $K(\, x\, )$ is a certain QED kernel \cite{Bouchiat} given by \ba
K(\, x\, ) & =&  x^2 \left(1 - \frac{x^2}{2} \right ) + (\, 1 + x\, )^2 \left(1 + \frac{1}{x^2} \right ) \left(\ln (\,  1 + x\, ) - x  + \frac{x^2}{2} \right ) \nn \\   & + & \frac{1+x}{1-x} x^2 \ln x, \nn 
\ea and $x(\, s\, ) = (\, 1 - \beta_\mu\, ) / (\, 1 + \beta_\mu\, ) $ and $\beta_\mu = \sqrt{ 1 - 4 m_\mu^2 / s } $.  We noticed that current literature imposes the lower limit of integration  $s_{min} = 4 m_{\pi}^{2}$ in Eq.~\ref{dispint}, where $m_{\pi}$ is the charged pion mass.  The threshold comes from assuming the lowest energy hadronic final state is a pion-anti pion pair at rest. 

Eq. \ref{form} originates in dispersion relations for complex analytic functions. Setting $\sqrt{s_{min}}=2 m_{\pi} \sim 279$ MeV is an error which neglects states below threshold.  The magnitude and sign of the $a_{\mu}$ discrepancy is a shortfall of about 28 $\times 10^{-10}$ units, relative to some 690 $\times 10^{-10}$ units integrating $s>4 m_{\pi}^{2}$, suggesting that neglected states might account for the discrepancy.  Both continuum and bound states of discrete energy must be taken into account.  

Among continuum states there are several contributions, but all appear small. Pions are unstable particles for which ``anomalous'' thresholds exist, a tiny effect. Next, the production of $\pi^{0}$ plus photon (\, $\gamma$\, ) starts at $\sqrt{s} \geq 135$ MeV.  The famous ``axial anomaly'' reasonably predicts the $\pi^{0} \ra \gamma \gamma$ coupling. This channel has been previously noted in the region well above threshold, and for 500 MeV $<\sqrt{s} < $1.8 GeV it contributes \cite{Davier2002}  $0.93\ 10^{-10}$ units to $a_\mu^{had, LO}$.  Simple estimates find the sub-threshold continuum ``anomalies'' too small for our ``anomaly. '' 

We turn to neglected bound states.  The Coulomb interaction will form pionic ``atoms'' bound in the long distance region.  The bound states contributing to vacuum polarization have quantum numbers $J^{PC}=1^{--}$, where $J$ is the spin, $P$ the parity and $C$ is charge conjugation.  Both isospin $I=0, \, 1$ are allowed.  A spin-1 ``$P$-wave'' pionic state contributing directly to vacuum polarization undoubtedly exists. In order to establish a discrepancy between conventional physics and calculations, one must first resolve the contribution of the omitted state(\, s\, ). 

It may appear simple to calculate the spin-1 state's contribution to $a_{\mu}$.  There are several ways to make an estimate.  One method approaches the problem perturbatively, and extrapolates the Coulomb atomic wave function into the core region. Following the treatment outlined in Ref. \cite{Jauch}, the decay width from a $P-$ wave state of pointlike, spinless constituents is given by \ba \Gamma (\, spin-1 \ra e^+ e^-\, ) = 8 \alpha^2 \frac{|R'(\, 0\, )|^2}{m_{\pi_{2}}^4} \label{width}. \ea  Here $\a$ is the fine structure constant and $R'(\, 0\, )$ is the derivative of the Schr\"odinger radial wave function at the origin. There is an important issue of what might be meant by the ``Schr\"odinger radial wave function at the origin'' for pions, which have finite size, but we continue.  In the Coulomb model $R'(\, 0\, ) \sim (\, \a m_{\pi}\, )^{5/2}$ (\, see below\, ), and the coupling to $e^{+}e^{-}$ is far too small to matter.  For another estimate recall that the $\rho$ meson mixes with the photon at the order of $\a$.  The vector, isovector bound state of pions must also mix with the $\rho$ and $\gamma$, giving no reason to forbid mixing of order $1\%$ or larger, especially given the large $\rho \pi \pi$ coupling. This estimate then produces a contribution too large to explain the data for $a_{\mu}$.   

Why do basic estimates produce such large variation?  In Eq. \ref{width} we see $(\, \a^{5/2}\, )^{2} \sim 10^{-11}$ rules the perturbative Coulomb model, which has very little wave function near the origin.  Strong interactions and $\rho$ mixing, however, produce all their effects in the 1 Fm region of the origin.  Since the Coulomb model predicts nearly ``zero,'' then neglecting effects in the strongly interacting core of the state produces an enormous relative error in the contribution to $a_{\mu}$.  The only way one might confidently use the ``Coulomb core'' model would be to know in advance that strong interaction effects and $\rho$ mixing are {\it even smaller} than the electromagnetic binding effects, which we find absurd.  
%
%

%

Since strong interactions of hadrons at short distances are a non-perturbative problem it would be unwise to proceed on model estimates.  The physical, strongly interacting mixed state needs a particular name: We call the spin-1, isospin-1 state {\it pi-rhonium}, symbol $\pi_{2/\rho}$.

For a model-independent approach\footnote{Nothing in this approach hinges on the precise value of binding energy, which needs to be found experimentally.} we investigated the coupling needed for pi-rhonium to be relevant to $a_{\mu}^{had, LO}$.  It is safe to assume the width $\Gamma(\, \pi_{2/\rho} \rightarrow e^+e^-\, )$ is small and treat its effects as an isolated spike (\, delta function\, ) in the integration. A short calculation gives \ba  
a_\mu^{\pi_{2/\rho}} = \frac{3}{\pi} \frac{K(\, m_{\pi_{2/\rho}}^2\, )}{m_{\pi_{2/\rho}}} \Gamma(\, \pi_{2/\rho} \rightarrow e^+e^-\, ) . \nn \ea  If the state accounts for the full discrepancy indicated by Eq. \ref {disc} we then have $\Gamma(\, \pi_{2/\rho} \rightarrow e^+e^-\, )=28$ eV. 

It is interesting that the energy region to detect $\pi_{2/\rho}$ appears never to have been searched. The energies involved seem to have been {\it too low} to be ``interesting.''  The first $e^{+}e^{-}$ colliding beams at Frascati \cite{Bernardini} apparently searched well above pion pair threshold to assure data would exist.  Data available \cite{data} on $e^{+}e^{-} \ra \pi \pi$ begins at energies well above threshold: the high precision CMD-2 experiment \cite{CMD2} recently reported on $\sqrt{s} >$ 370 MeV.  

\begin{figure}
\begin{center}

\includegraphics[width=4.5in]{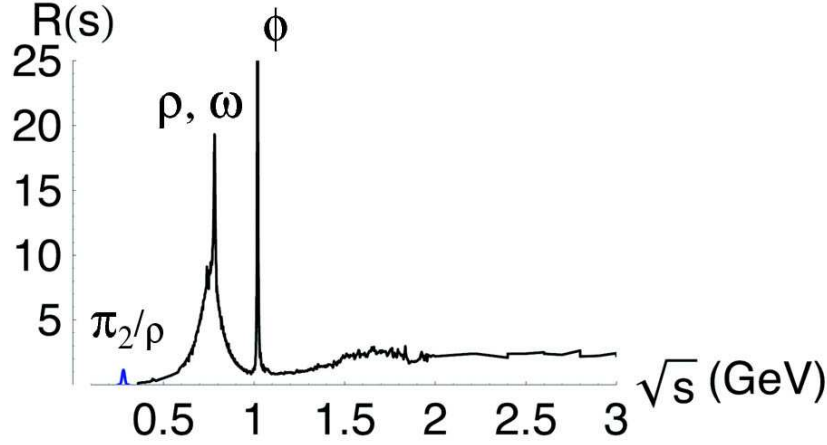}

\caption{\small Data\cite{data} for the ratio $R(\, s\, ) = \sigma(\, e^{+}e^{-} ~\ra hadrons\, )/\sigma(\, e^{+}e^{-} \ra \mu^{+}\mu^{-}\, )$ versus center of mass energy $\sqrt{s}$ with peaks due to the $\rho, \omega$ and $\phi$ mesons.  The contribution of the state $\pi_{2/\rho}$ is predicted to be the small bump below threshold to produce pions. The $\pi_{2/\rho}$ width has been convolved with 10 MeV energy resolution.}
\label{fig:Rfig}
\end{center}
\end{figure}

The information can be reversed to predict the effects on $\sigma(\, e^{+}e^{-} \ra hadrons\, ).$  Fig. \ref{fig:Rfig} shows a tiny bump in $R(s)$ that should be observed at energies just below $2 m_{\pi}$.  The observable effect in $a_{\mu}^{had, LO}$ is non-negligible because the convolution with $K(\, s\, )/s$ (\, Eq. \ref{form}\, ) greatly exaggerates the small $s$ region. The width of the bump shown includes a 10 MeV figure for the energy resolution of the beam and detectors, based on performance of modern $e^{+}e^{-}$ colliders operating in the several GeV region.  
 
Assuming energy resolution scales in proportion to system energy, the calculation indicates that devices operating with current technology in the 200-300 MeV region can discover $\pi_{2/\rho}$.  Signals include the $e^+e^-$ peak, and a $\mu^{+} \mu^{-}$ channel we predict with $\Gamma(\, \pi_{2/\rho} \rightarrow \mu^+\mu^-\, ) \sim 0.6 \, \Gamma(\, \pi_{2/\rho} \rightarrow e^+e^-\, )$, based on the phase space.  Strong decays will also give signals discussed below.  Other experiments using rare particle decays, or hadronic beams on fixed targets are also attractive: the spin-1 nature of the state gives well-established angular distributions to decay products that provide distinctive signals.

This concludes the ``IF...THEN'' logic of our analysis, whose central elements do not depend on any model at all.  For example if 35\% of the $g-2$ discrepancy comes from hadronic light-by-light scattering, the predicted width will be revised.  Our analysis also changes the role of ``new physics'' contributions.  At most $a_{\mu}$ constrains the sum of $\pi_{2/\rho}$ and new-physics contributions, opening the door for a wider range of new physics than constrained to obtain $a_{\mu}$ exactly with fine-tuned parameters. (\, There is no ``fine-tuning'' of $\pi_{2/\rho}$ properties as the remainder, because the observed discrepancy of $a_{\mu}$ is defined by what is experimentally measured.\, )  IF the state $\pi_{2/\rho}$ is observed in $e^{+}e^{-}$ channels within error bars of the width estimated, THEN the anomalous moment problem is solved. 

\medskip

2. The model-independent analysis is complemented by exploring reasonable models in more detail.  It is interesting to see what strong interaction phenomenology may give, while keeping in mind that experimental resolution is invariably the only resolution of non-perturbative strong interaction questions. 

{\it Wave Function Formalism:}  Suppose one works with Schr\"odinger wave function formalism.  Most previous studies of the $S$-wave pionic atom, called ``pionium'', adopt this approach.  

The literature on pionium is very old, and comes in two varieties.  There is a long line of research on the bound state physics. We need the effects of the short-distance strong interactions on the {\it derivative} of wave functions.  Suebka's solution to the $S$-wave is illustrative\cite{suebka}. It shows a jump of relative order unity in a distance of order $1/m_{\pi}$. The average derivative $\Delta \psi/\Delta r$ changes by a factor of about $1/\a \sim 10^{2}$.  This reiterates that pi-rhonium's contribution to $a_{\mu}$ is exquisitely sensitive to
the model of the wave function due to three effects: first, the characteristic size $\Delta x$ changes enormously from Bohr radius-level to Fm-scale level (\, a factor of order 390\, );  second, the change of the $P-wave$ is compounded by the kinematic dependence to the 5/2 power, and third the wave function is squared to get the coupling contributing to $a_{\mu}$.  Finally the wave function of a potential model cannot be disentangled from how the wave function is used in mixing, discussed shortly.

\begin{figure}
\begin{center}

\includegraphics[width=4in]{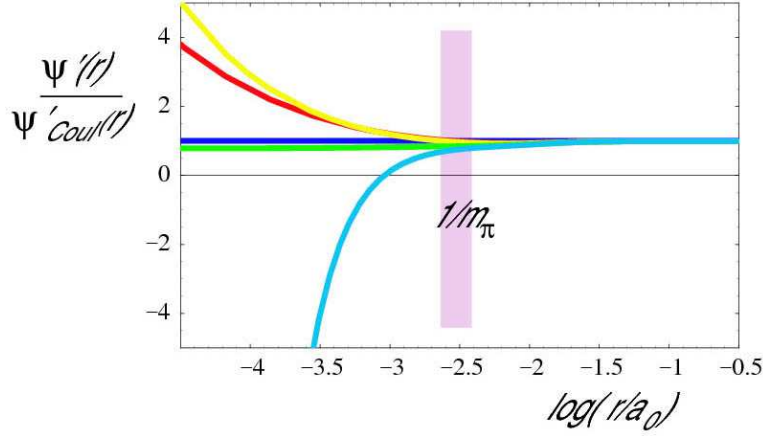}

\caption{Illustration of model sensitivity in a toy model: the ratio of space derivatives of Klein-Gordon-Yukawa $P$-wave functions compared to the non-relativistic Coulomb $P$-wave function. Couplings $g_{\rho \pi \pi}$ equals 0 (blue), 3.53 (red), 3.74 (yellow), 4.21 (green), 4.33 (light blue).  Fat vertical line shows the region $r \sim 1/m_{\pi}$. }
\label{fig:waves}
\end{center}
\end{figure}

We could find no discussion of $P$-wave $I=1$ pionium in the literature. This is
perhaps because the mixing problem is hard, and pionium work focuses on the scalar state decaying to $\pi_{0} \pi_{0}$. The $P$-wave non-relativistic model has a centrifugal barrier, but the effective centrifugal barrier of strongly interacting systems depends on the potential model. In the Klein-Gordon Coulomb problem, which can be solved exactly\cite{bethe}, the non-relativistic centrifugal barrier $l(\, l+1\, )/r^{2} \ra l(\, l+1\, )/r^{2}+V^{2}(\, r\, )+{\mathcal O}(\, V\, )$ where $l$ is the orbital angular momentum and $V(\, r\, )$ is the potential of a vector interaction. This is a textbook consequence of gauge invariant vector couplings. We found it interesting to explore numerical solutions  (\, Fig. \ref{fig:waves}\, ) of a toy Coulomb plus Yukawa model with coupling $\a_{g} = g_{\rho \pi \pi}^{2}/4 \pi$.  Range $1/m_{\pi}$ minimizes parameters and serves as something representing the pion size.  It is well known\cite{bethe,relativistic,sucher} that sufficiently strong potentials overcome the non-relativistic centrifugal barrier.  Solutions become unstable at coupling $g_{\rho \pi \pi}\geq 4.2, \, \a_{g} \geq 1.4$.  (\, In fact the $P$-wave $R'(\, 0\, )$ diverges when literally evaluated at zero {\it regardless of coupling strength}.\, ) A large literature on relativistic bound states of Klein-Gordon and Dirac type explores many variations, including the interesting fact\cite{sucher} that regulating the $1/r$ singularities with smeared out densities does not remove strong-coupling instability.  Softening of the centrifugal barrier is not only a feature of the Klein-Gordon system, it is seen in the Bethe-Salpeter formalism\cite{Bethe-Salpeter} and in the spinless Salpeter equation\cite{spinlessSalp}  that uses Hamiltonian $H =\sqrt{ p^{2}+m^{2}}+V$.  One can evidently fit a value of $R'(\, 0\, )$ with a defensible range of $g_{\rho \pi \pi}$.  Yet the importance of the anomalous moment problem is too high to let $\pi \pi$ interactions models decide it one way or the other. 
 
Are there experimental constraints from pionium? Much pionium research fixes the isosinglet $S$-wave function to the Coulomb one, and seeks to calculate the pionium lifetime via annihilation $\pi^{+} \pi^{-} \ra \pi^{0} \pi^{0}$. Note that mixing for the scalar isoscalar (\, $S$-wave\, ) state is a much different issue than the vector-isovector state we discuss.  There is no model independent way to go from the $S$-wave to the $P$-wave.  Recently the DIRAC experiment\cite{Dirac} reported observation of pionium in experiments sending particle beams through metal foils.   The experiment did not measure the spin or isospin of the state.  What was actually measured were certain correlations of longitudinal pion momenta.  The correlations were interpreted as a lifetime after subtracting a Monte Carlo model of propagation and backgrounds.  If the $P$-wave couples substantially, as $a_{\mu}$ indicates, the interpretation of what was observed might be revised.  There are too many assumptions and too many steps for the limited observations made so far to say anything about the $\pi_{2/\rho}$ contribution.

{\it Mixing Models:}  Whether or not one might estimate the wave function of the spin-1 pionic state on its own, quantum mechanical mixing creates new complexity.  We explored a simple two state mixing model with off-diagonal mixing terms.  We mix isovector states and ignore the $\omega$ meson for simplicity.  

The bare basis is denoted $\ket{\tilde \pi_{2}}, \, \ket{\tilde \rho}$, whose linear combinations make the physical $\ket{  \pi_{2/\rho}}, \, \ket{  \rho}$: \ba  \left(\, \begin{array}{c}\ket{  \pi_{2/\rho}} \\ \ket{ \rho}\end{array}\right )= \left(\, \begin{array}{cc} cos\beta &  sin\beta \\ -sin\beta & cos\beta  \end{array}\right ) \left(\, \begin{array}{c}\ket{ \tilde  \pi_{2}} \\ \ket{ \tilde \rho}\end{array}\right ).  \nn \ea

The Hamiltonian matrix elements in the bare basis are
\ba
H_{int} =  \left(\begin{array}{cc} m_{\tilde\pi_{2}} & m_{ix} \\ m_{ix} & m_{\tilde\rho} \end{array} \right ).
\nn \ea
Here $m_{\tilde{\rho}}$ and $m_{{\tilde\pi_{2}}}$ are bare masses, and the mass mixing term is named $m_{ix}$.  Diagonalization gives the mixing angle $\beta$, \ba
\tan \beta = \frac{1}{2 m_{ix}} (\, \, m_{\tilde{\rho}} - m_{{\tilde\pi_{2}}} - \sqrt{(\, m_{\tilde{\rho}} - m_{{\tilde\pi_{2}}}\, )^2 + 4 m_{ix}^2} \, \, ).
\nn \ea
To proceed we must fix three parameters $m_{\tilde{\rho}}$, $m_{{\tilde\pi_{2}}}$, and $m_{ix}^2$. Eigenvalues yield the masses of the physical states, namely $m_{\pi_{2/\rho}}$ and $m_{\rho}$. Setting the physical mass $m_\rho = 775.5$ MeV fixes one parameter.  

Continuing, we need reference values for the bare states. In quantum field theory the calculation will go through a Bethe-Salpeter wave function and irreducible kernels connecting the pionic to photon degrees of freedom. Rather than clutter the calculation with definitions, we express these concepts {\it in units of} the non-relativistic Schr\"odinger model with Coulomb potential.  Notice this is the smallest wave function we might use.  No particular belief in a Coulomb core is required, because other wave functions are accomodated by re-scaling. For Hydrogen-like atoms the binding energy is given by
$E_{b}^{H} = - \mu \alpha^2/(\, 2 n^2\, )$, where $n\ra 2$ is the principal quantum number in this case.  This predicts a reference value for bare $m_{ \tilde\pi_{2} }$, fixing a second parameter.  

The hydrogenic wave function for the $2P$ state is 
\ba \label{wavefun}
R_{{\tilde\pi_{2}}}(\, r\, )=\frac{1}{\sqrt{24}} \frac{r}{a_0^{5/2}} e^{-r/2a_0},
\nn \ea
with Bohr radius $a_{0} =2/(\, \a m_{\pi}\, )$ and the reduced mass $\mu = m_\pi / 2$.  The normalization is $\int_0^\infty \d r r^2 |R(\, r\, )|^2 = 1$.  The derivative of the bare $P$-wave pionium wave function at the origin is then 
$ R_{{\tilde\pi_{2}}}'(\, 0\, ) = \frac{1}{\sqrt{768}} \alpha^{5/2} m_\pi^{5/2}.$  For consistency of this toy calculation we use the same framework for the $e^{+}e^{-}$ channels of the $\rho$, with width $\Gamma(\, \rho \ra e^+e^-\, )  = 4.70\ 10^{-5}\ \Gamma_\rho=6.8$ keV.  

Finally we use the mixing of wave functions to calculate the width $\Gamma(\, \pi_{2/\rho} \rightarrow e^+e^-\, ) \sim 28$ eV.  It fixes the last parameter, so none remain free.  Since the bare pionium and bare $\rho$ are orthogonal states, we have
\ba
|R_{\pi_{2/\rho}}'(\, 0\, )|^2 & = & \cos^2 \beta |R_{{\tilde\pi_{2}}}'(\, 0\, )|^2 + \sin^2 \beta |R_{\tilde{\rho}}'(\, 0\, )|^2 ;\nonumber \\
|R_\rho'(\, 0\, )|^2 & = & \sin^2 \beta |R_{{\tilde\pi_{2}}}'(\, 0\, )|^2 + \cos^2 \beta |R_{\tilde{\rho}}'(\, 0\, )|^2. \nonumber
\ea 
Solving the equations gives 
\ba
\sin \beta = 0.81 \% ; \quad \beta = 0.46^\circ. \nonumber 
\ea
The wave function of $\pi_{2/\rho}$ {\it need only be mixed by order the fine structure constant} $\a =1/137$, which we find satisfactory.  It is just the same order of magnitude of mixing one might guess on general grounds. 

With no freedom left the physical mass is $m_{\pi_{2/\rho}}  = 279.11 {MeV}$. Several digit accuracy is reported for completeness and to exhibit small binding effects.  The estimated binding energy is about 32 keV, or 65 times the ``Bohr-atom'' estimate.  The mixing interaction energy $m_{ix} =3.96 \,$ MeV.  This coupling at order $\sim \a \times  2 m_{\pi}$ seems at the lower limit of what is reasonable.  For one thing, finer discrimination would need control of electromagnetic (\, $\pi^{+} - \pi^{0}$\, ) mass differences. 

These calculations are done in units of the Coulomb core for the bare $P$-wave. Keeping the width of $\pi_{2/\rho}$ fixed by data, larger core wave functions will decrease the mixing angle correspondingly.  It should be clear the idea of wave functions of pions in the $\rho$ is somewhat naive, and that much more information about the short-distance nature of the $\rho$ is needed to believe wave function or mixing calculations. The wave function and the mixing ambiguities are complementary ``microscopes'' for looking into the interior of the state.  In the core we believe that quarks are relevant, and they suggest an interesting model summarizing model requirements.

In the quark basis production of pi-rhonium goes through a pointlike photon vertex where quark chirality is conserved for light $u, \, d$ quarks with masses of a few MeV.  The quark-antiquark pair when produced have spins aligned in an orbital $S$-wave.  That is, the quark wave function literally at the origin is just like that of a $\rho$.  This is a sure fact of short-distance QCD perturbation theory.  Since the quarks are in an $S$-wave, no centrifugal barrier exists in the interior of the 1 Fm confining distance scale.  

If the mass is as predicted, the $\pi_{2/\rho}$ may be lighter than the $S$-wave, and therefore the second lightest hadron, coming right after the pion triplet.  This causes a minor puzzle.  It is sometimes thought that the lowest state of any bound system is an $S$-wave.  The precise statement is that the ground state must have no nodes. Feynman\cite{Feynman} gives a simple proof that does not depend on the Hamiltonian. The overall {\it quark-model} wave function of $space \otimes spin$ is then a mixture of $S$ and $P$ waves and has no nodes.  As the quarks emerge into the Bohr radius regime, they are dressed into pions, converting quark spin to orbital angular momentum of the pions.  When pions return to short distance they re-convert orbital angular momentum to quark spin.  Consider the same physical picture in {\it pion basis}.  Let $v$ be the velocity of the pions, and estimate the radius $r_{min}$ where pions separate to develop one unit of orbital angular momentum: \ba r_{min} \times m_{\pi}v \sim \hbar .\nn \ea  
If the pions are non-relativistic $v << 1$, and $r_{min} >> \hbar/ m_{\pi}$.  However this contradicts confinement, from which we know pions form in the region $r \lesssim \hbar/m_{\pi}$. This tells us one cannot extrapolate any non-relativistic $P$-wave model to Fm distances.  The consequences of quark spin-orbital effects, in fact, are that neither the $\pi_{2/\rho}$ core nor the $\rho$ meson can sensibly be conceived with pions. Yet pions are perfectly consistent far away from the core.  We can calculate the velocity $v_{0} \sim \a$ from Coulomb physics.  Then we have $r_{min}  \sim \hbar/v_{0}m_{\pi} \sim 1/(\, \a m_{\pi}\, ) $, which is just the Bohr radius $r_{min}\sim a_{0}$.  If the physical picture seems a bit visionary, the $\rho$ meson accomplishes an identical conversion of quark spin to pion orbital angular momentum in every strong decay $\rho \ra \pi \pi$.  This is why we know it applies. The strong core of $\pi_{2/\rho}$ is a tiny effect.  It has an overall probability $sin^{2}\b \lesssim 10^{-4}$.  Yet $a_{\mu}$ is a special observable that is highly sensitive to small probability due to the close coincidence of the muon and pion masses. 

There are also time scale issues to check.   Confinement shows that quarks convert to pions in a time of order 1 Fm/$c$, which is fast compared to the bound state oscillation time scales. The bound state ``orbit time'' $T_{orbit}$, is semi-classically estimated from $T_{orbit}\sim a_{*}/v_{*}$, where $a_{*}$ is the largest spatial size of the state and $v_{*}$ a characteristic velocity.  Semi-classical consistency needs $T_{orbit} \lesssim \tau$, where $\tau$ is the lifetime of the state.  We cannot calculate the lifetime, but we observe the strong decay $\pi_{2/\rho} \ra \pi^{0}\gamma $ is allowed in the core.  We create a rough estimate by taking $ \Gamma(\, \pi_{2/\rho} \ra e^{+}e^{-}\, ) = $ 28 eV and rescaling it by proportions found in the $\rho$ meson.  By this $\Gamma_{\pi_{2/\rho}, \, tot} \sim   28 \, eV  \Gamma_{\rho\ra \pi \gamma}/\Gamma(\, \rho \ra e^{+}e^{-}\, ) \times ps,$ where $ps$ is a calculable phase space ratio. Evaluating the formulas gives $T_{orb}/T_{decay} \sim {\cal O}(\, 10^{-3}\, )$, which is very acceptable. 

The core amplitude of $\pi_{2/\rho}$, namely the region contributing to the muon's anomalous moment, must be a small strongly interacting object that acts as a new hadron. Unless deeply bound the mass of $\pi_{2/\rho}$ scales with $m_{\pi}$, and vanishes in the chiral limit.  It is very interesting to ask if physics of the pion as a Goldstone boson predicts any massless vector particles.  The non-linear sigma model uses a derivative expansion of the couplings of unitary fields $U$. Within that context, massless vector mesons have long existed in gauge-invariant low-energy derivative expansion taking the form  \ba L  =1/2 \, tr (\, \p_{\mu} U -V_{\mu}\, )  (\, \p_{\mu} U -V_{\mu}\, )^{\dagger} \, ) -tr(\,  F^{\mu \nu}F_{\mu \nu}\, ) + ...\label{model} \ea  Here $tr$ indicates the trace, $V_{\mu}$ is the vector combination of $U\p_{\mu}U^{\dagger}$ and  $F_{\mu \nu} $ is its particular gauge field strength.  Traditionally the gauge-like fields are interpreted as the $\rho$ meson.  The need to give the $\rho$ meson a mass then forces {\it ad hoc} breaking of the  symmetries of the theory. Yet we have $\pi_{2/\rho}$ a candidate massless meson in the chiral limit.  Was it overlooked twice? It is not necessary that $\pi_{2/\rho}$ remain in the spectrum as electromagnetism is turned off. The sigma model is so restrictive that a vector state couples as Eq. \ref{model} predicts as electromagnetism is turned {\it on}.  Certainly $\pi_{2/\rho}$ will mix with the effective vector fields of the chiral theory.  And how much does it mix? This is a fascinating avenue, with the implications that the couplings of soft pions to $\pi_{2/\rho}$ might be theoretically interesting.
  
\medskip

3.  {\it To summarize: }   A light pionic state with $J^{PC}=1^{--}$ is known to exist and was long overlooked. Its contribution to $a_{\mu}$ is significantly enhanced due to kinematic factors in the basic formula of Eq. \ref{form}.  The state stands to explain the muon magnetic moment anomaly if the $\rho$ mixing amplitude is of order $1 \%$.  This is a modest and credible requirement, given the uncertainties of strong interaction phenomenology.  We have made more than a good faith effort to check reasonableness of the partial width and its role in $a_{\mu}^{had, \, LO}$.  Calculations are non-perturbative and model dependent, and far more work would be needed to claim the physical state has the width needed.  Conversely it would be irresponsible to dismiss the contribution of $\pi_{2/\rho}$ on the basis of wave function or mixing models at their current level of development. 

The burden of proof is reversed in our view.  If there is truly a discrepancy between QED and experiment, it seems necessary for proponents of the discrepancy to resolve the pi-rhonium contribution.  The door is opened for new physics to explain {\it part} of the discrepancy without obligation to explain {\it all} the discrepancy.  The most productive resolution will come by comparing experiment to experiment.  If pi-rhonium accounts for the $a_{\mu}$ discrepancy, there are observable consequences in the experimentally observable couplings of $\pi_{2/\rho} \ra e^{+}e^{-}, \, \ra \mu^{+}\mu^{-}, \, \ra \pi^{0}\gamma \ra  \gamma \gamma \gamma$, and similar channels. 

\medskip

{\it Acknowledgments:} Research supported in part under DOE Grant Number
DE-FG02-04ER14308. We thank Danny Marfatia, Doug McKay, Sandip Pakvasa, Graham Wilson, Alice Bean, Vernon Barger, and an anonymous referee for comments.

\end{document}